# A Monte Carlo Calculation of Atmospheric Muon and Neutrino Fluxes[†]

M. Thunman[a*], G. Ingelman[a] and P. Gondolo[b]

[a]Department of Radiations Sciences, Uppsala University, Box 535, S-75121 Uppsala, Sweden

[b]LPTHE, Université de Paris VI-VII, 2 place Jussieu, F-75251 Paris Cedex05, France

Production of muons and neutrinos in cosmic ray interactions with the atmosphere has been investigated with a cascade simulation program based on Lund Monte Carlo programs. The resulting 'conventional' muon and neutrino fluxes (from $\pi, K$ decays) agree well with earlier calculations, whereas the improved charm particle treatment used in this study gives significantly lower 'prompt' fluxes compared to earlier estimates. This implies better prospects for detecting very high energy neutrinos from cosmic sources.

## 1. Introduction

The flux of muons and neutrinos at the earth has an important contribution from decays of particles produced through the interaction of cosmic ray particles in the atmosphere. This has an interest in its own right, since it reflects primary interactions at energies that can by far exceed the highest available accelerator energies. It is also a background in studies of neutrinos from cosmic sources as attempted in large neutrino telescopes, such as AMANDA, BAIKAL, DUMAND and NESTOR. Here we give a short report of a comprehensive study of muon and neutrino production in cosmic ray proton interactions with nuclei in the atmosphere using detailed Monte Carlo simulations [1,2].

The atmospheric muons and neutrino fluxes are dominated by 'conventional' sources, i.e. decays of relatively long-lived particles such as $\pi$ and $K$ mesons. This is well understood from earlier studies [3,4] and confirmed by our investigations [1,2]. With increasing energy, the probability increases that such particles interact before decaying. This implies that even a small fraction of short-lived particles can give the dominant contribution to the high energy tail of the muon and neutrino fluxes. These 'prompt' muons and neutrinos arise through semi-leptonic decays of hadrons containing heavy quarks, most notably charm.

Previous estimates of the flux of prompt muons and neutrinos from charm [5–7] vary by a few orders of magnitude (cf. Fig. 3). This is due to the different models used to calculate the charm hadron cross section and energy spectra. The models are usually constrained or fitted to experimental data at accelerator energies and then extrapolated to the higher energies needed in this context. Obviously, the extrapolation can only be trustworthy if based on a sound physical model. The main new contribution of our study is in this context, where we apply state-of-the-art models to simulate charm particle production in high energy hadron-hadron interactions.

## 2. The cascade algorithm

The prompt flux is essentially independent of the zenith angle due to the negligible probability that the very short-lived particles interact before decaying. This flux is therefore relatively more important in the vertical direction where the conventional flux is lower. The fluxes are therefore calculated for the vertical direction which has the advantage that a simple exponential model for the density of the atmosphere can be used, i.e. the atmospheric depth is

$$X[\mathrm{g/cm}^2] = 1300\, e^{-h[\mathrm{km}]/6.4} \qquad (1)$$

The parameters are chosen to give a good description of the atmosphere at altitudes of 10–40 km where the interactions typically take place.

---

*rapporteur, thunman@tsl.uu.se

[†] 

The simulation algorithm is constructed as follows [1,2]. The energy of a cosmic ray proton is chosen from the form $\Phi(E) \sim E^{-\gamma-1}$, with $\gamma = 1.7$ for $E$ below $5 \cdot 10^6$ GeV and $\gamma = 2$ above. A proton-nucleon interaction is then generated with PYTHIA [8] resulting in a complete final state of particles. Cross section and mean free path are scaled according to the composition of the atmosphere, and an interaction height is chosen. The produced particles are traced through the atmosphere until they either decay or interact as given by simulated decay lengths and interaction lengths. Particle decays are fully simulated with daughter particle momenta. In case of interactions, the interacting particle is regenerated in the same direction but with degraded energy corresponding to the appropriate leading particle spectrum. The procedure is repeated until all particles have decayed, hit the ground or their energy fallen below a minimum of $10^2$ GeV. Finally, muons and neutrinos are counted with appropriate weights to obtain their energy spectra.

The simplification that secondary interactions are not simulated in full detail is justified since only leading particles will contribute significantly. Other particles with lower energy give a negligible flux compared to the flux of primary particles at the corresponding energy. Furthermore, proton-nucleus and meson-nucleus collisions are very similar except for the leading particle.

To obtain the flux of conventional muons and neutrinos, an inclusive event sample is simulated using PYTHIA in a mode generating minimum bias proton-proton interactions (including diffractive scattering). The particle production results from Lund model hadronisation of colour string fields between partons scattered in semi-soft QCD interactions. The prompt flux is obtained from charm particles arising from the hadronisation of charm quarks produced in the processes $gg \to c\bar{c}$ and $q\bar{q} \to c\bar{c}$ as calculated with leading order perturbative QCD (pQCD) matrix elements.

## 3. Numerical results

From the extensive studies in [1,2] we show in Fig. 1 the main results in terms of the calculated fluxes of muons and neutrinos. The prompt contribution from charmed particles dominates at high energies. This contribution is subdivided on the different charmed particles in Fig. 2, demonstrating the dominance of the $D^{\pm,0}$ mesons.

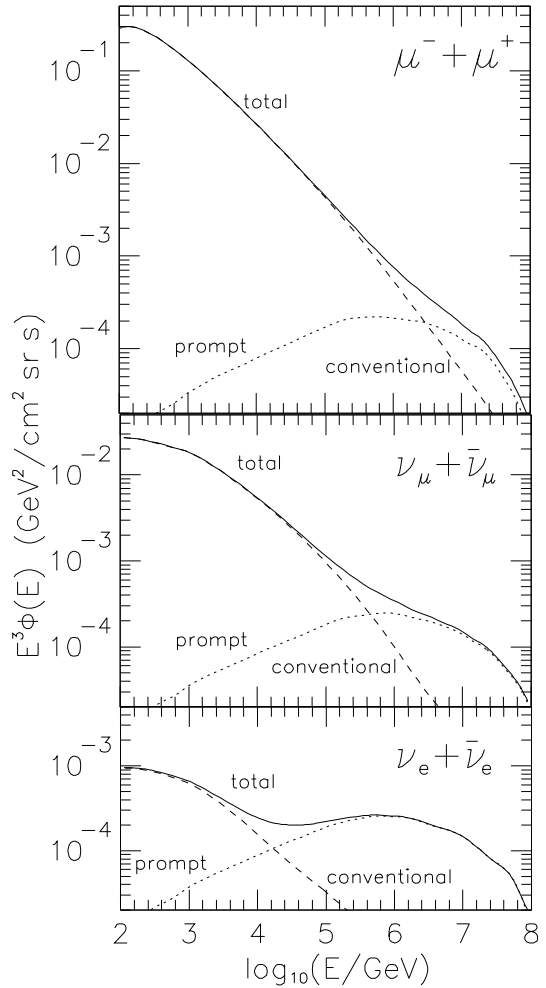

Figure 1. Vertical fluxes ($E^3$-weighted energy spectra) of muons and neutrinos from conventional ($\pi, K$ decays) and prompt (charm decays) sources and their sum as obtained from simulated cosmic ray interactions in the atmosphere.

The results for the inclusive prompt and conventional fluxes can be parametrised as (cf. the primary flux)

$$\phi(E) = \begin{cases} N_0 \, E^{-\gamma-1}/(1 + A\,E) & E < E_0 \\ N_0' \, E^{-\gamma'-1}/(1 + A'\,E) & E > E_0 \end{cases} \quad (2)$$

Table 1
Values of parameters in eq. 2 obtained from fits to the Monte Carlo results in Fig. 1.

|  | $N_0$ | $\gamma$ | $A$ | $E_0$ | $\gamma'$ | $A'$ | $N_0'$ |
|---|---|---|---|---|---|---|---|
| Conv. $\mu^- + \mu^+$ | $1.3 \cdot 10^{-1}$ | 1.77 | $4.4 \cdot 10^{-3}$ | $2.7 \cdot 10^5$ | 2.40 | $9.2 \cdot 10^{-6}$ | $4.6 \cdot 10^{-1}$ |
| Prompt $\mu^- + \mu^+$ | $2.4 \cdot 10^{-6}$ | 1.63 | $6.5 \cdot 10^{-7}$ | $3.2 \cdot 10^7$ | 2.64 | $6.2 \cdot 10^{-8}$ | $1.2 \cdot 10^1$ |
| Conv. $\nu_\mu + \bar\nu_\mu$ | $8.5 \cdot 10^{-3}$ | 1.55 | $9.1 \cdot 10^{-3}$ | $7.9 \cdot 10^3$ | 2.69 | $1.0 \cdot 10^{-6}$ | $3.3 \cdot 10^{-4}$ |
| Prompt $\nu_\mu + \bar\nu_\mu$ | $2.5 \cdot 10^{-6}$ | 1.63 | $6.5 \cdot 10^{-7}$ | $4.4 \cdot 10^7$ | 2.78 | $1.3 \cdot 10^{-8}$ | $8.0 \cdot 10^1$ |
| Conv. $\nu_e + \bar\nu_e$ | $3.8 \cdot 10^{-4}$ | 1.70 | $3.9 \cdot 10^{-3}$ | $1.3 \cdot 10^5$ | 2.32 | $4.6 \cdot 10^{-6}$ | $1.8 \cdot 10^{-3}$ |
| Prompt $\nu_e + \bar\nu_e$ | $2.5 \cdot 10^{-6}$ | 1.63 | $6.5 \cdot 10^{-7}$ | $4.1 \cdot 10^7$ | 2.87 | $6.0 \cdot 10^{-7}$ | $6.4 \cdot 10^3$ |

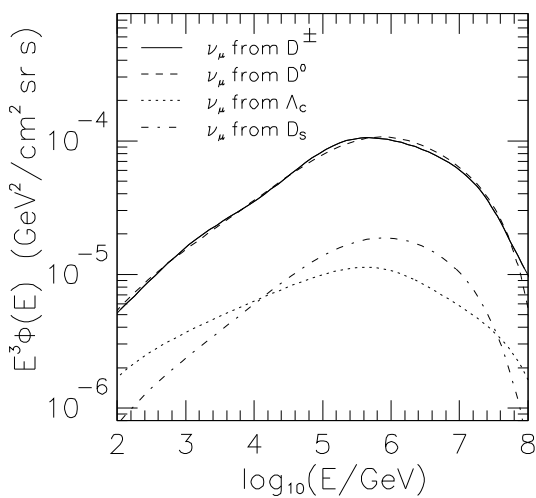

Figure 2. Flux of muon neutrinos and antineutrinos from decays of the different charmed particles.

distribution in longitudinal momentum fraction $x_F = p_\parallel/p_{max}$ (Feynman-$x$) of the charmed particles. Model I [6] uses a parametrised energy dependence of the cross section, normalised to some charm production data, and an $x_F$-spectrum of the form

$$dN/dx_F \sim (1 - x_F)^\alpha \qquad (3)$$

with $\alpha_D = 5$ and $\alpha_{\Lambda_c} = 0.4$ for $D$-mesons and $\Lambda_c$. Feynman scaling is thus assumed, i.e. $x_F$ is independent of the collision energy. Model II [7] simply takes charm as 10% of the total inelastic cross section and $x_F$ distributions of eq. (3) with $\alpha_D = 3$ and $\alpha_{\Lambda_c} = 1$. Model III [7] corresponds to charm quark production calculated with leading order pQCD matrix elements using relatively hard parton distributions.

with an accuracy of typically better than 10% using the fitted parameter values in Table 1.

### 4. Discussion

Whereas earlier calculations [3,4] agree well with our conventional muon and neutrino fluxes [1,2], previous estimates [6,7] of the prompt fluxes are up to orders of magnitude larger as illustrated in Fig. 3. (Prompt fluxes are direction independent up to $\sim 10^7$ GeV [1] and therefore directly comparable between horizontal and vertical directions.) These differences are due to different models for charm production, both regarding the energy dependence of the cross section and the

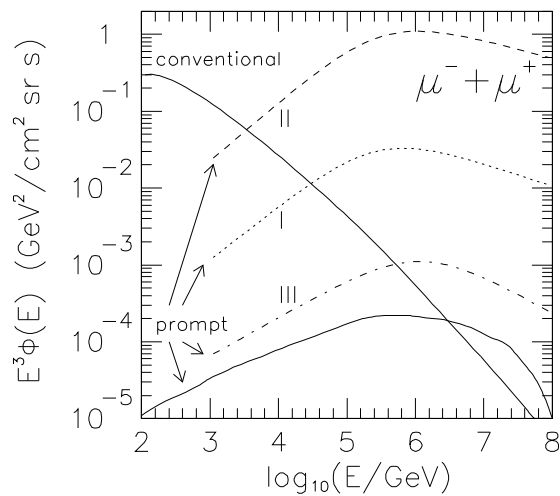

Figure 3. Prompt and conventional muon fluxes (solid lines) compared with earlier model calculations I–III of horizontal prompt fluxes (see text).

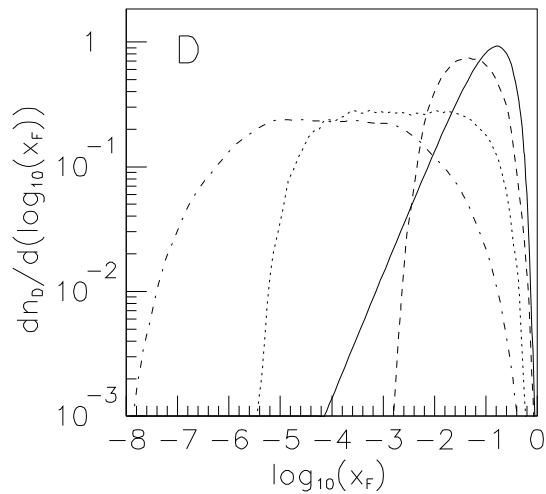 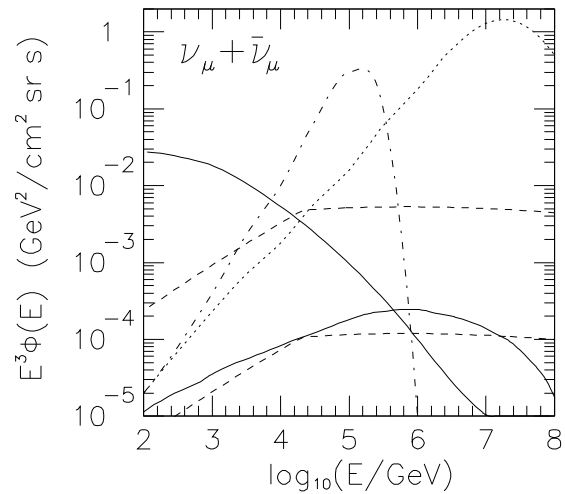

Figure 4. Distribution in $x_F$ of $D$-mesons in a model obeying Feynman scaling ($\alpha_D = 5$ in eq. (3), solid line), and from PYTHIA at proton beam energy $10^3$ GeV (dashed line), $10^6$ GeV (dotted line) and $10^9$ GeV (dash-dotted line).

Figure 5. Vertical fluxes of conventional and prompt atmospheric muon-neutrinos (solid lines) compared to astrophysical sources: the flux from cosmic ray interactions with the intra-galactic medium (dashed curves derived from [9]; upper curve apply for the direction towards the center of the galaxy, and lower curve orthogonal to the galactic plane) and the estimated diffuse fluxes from active galactic nuclei (dotted line [10], dash-dotted line [11]).

Whereas our model reproduces available data on charm production cross sections [2], not all of models I–III do. When extrapolating orders of magnitude in energy, the cross section differences are enhanced to factors of 10. A more important reason for our lower flux is, however, the strong breaking of Feynman scaling as demonstrated in Fig. 4. This arises from the dominance of charm production close to threshold in pQCD such that, with increasing energy, smaller momentum fractions $x$ of incoming partons will contribute (giving a sensitivity to parton density parametrisations). Together with the hadronisation mechanism, this dynamics imply smaller $x_F$ at higher collision energies.

Finally, the calculated atmospheric muon-neutrino flux is considered as a background in comparison to astrophysical sources, as illustrated in Fig. 5. The neutrino production in cosmic ray interactions with the galactic medium and, in particular, the diffuse neutrino flux from active galactic nuclei is estimated to be in excess of the atmospheric neutrino background for high energies and should therefore be possible to detect with large scale neutrino telescopes.